\documentclass[fleqn,usenatbib]{mnras}

\usepackage{newtxtext,newtxmath}
\usepackage{aas_macros}
\usepackage{float}

\usepackage[T1]{fontenc}

\DeclareRobustCommand{\VAN}[3]{#2}
\let\VANthebibliography\thebibliography
\def\thebibliography{\DeclareRobustCommand{\VAN}[3]{##3}\VANthebibliography}

\usepackage{graphicx}	
\usepackage{amsmath}	
\usepackage{caption}
\usepackage{subcaption}
\usepackage{graphicx}
\usepackage{subcaption}
\usepackage{mwe}
\usepackage{multirow}
\usepackage{array}
\usepackage{comment}
\usepackage{xcolor}
\newcolumntype{P}[1]{>{\centering\arraybackslash}p{#1}}

\title[Inferring binary black holes stellar progenitors]{Inferring binary black holes stellar progenitors with gravitational wave sources}

\author[S.~Mastrogiovanni]{
S.~Mastrogiovanni,$^{1,2}$\thanks{E-mail: smastro@oca.eu (SM)}
A.~Lamberts,$^{1,2}$
R.~Srinivasan,$^{1,2}$
T.~Bruel$,^{1,2}$
N.~Christensen$,^{2}$
\\
$^{1}$Université Côte d’Azur, Observatoire de la Côte d’Azur, CNRS, Laboratoire Lagrange, Bd de l’Observatoire, CS 34229, 06304 Nice cedex 4, France.\\
$^{2}$Artemis, Université Côte d’Azur, Observatoire de la Côte d’Azur, CNRS, F-06304 Nice, France\\
}

\date{Accepted XXX. Received YYY; in original form ZZZ}

\pubyear{2022}

\begin{document}
\label{firstpage}
\pagerange{\pageref{firstpage}--\pageref{lastpage}}
\maketitle

\begin{abstract}
With its last observing run, the LIGO, Virgo, and KAGRA collaboration has detected almost one hundred gravitational waves from compact binary coalescences. A common approach to studying the population properties of the observed binaries is to use phenomenological models to describe the spin, mass, and redshift distributions. More recently, with the aim of providing a clearer link to astrophysical processes forming the observed compact binaries coalescences, several authors have proposed to employ synthetic catalogs for population studies. 
In this paper, we review how to employ and interpret synthetic binary catalogs for gravitational-wave progenitors studies. We describe how to build multi-channel merger rates and describe their associated probabilities focusing on stellar progenitor properties. We introduce a method to quantify the match between the phenomenological reconstruction of merger rates with synthetic catalogs. We detail the implementation of synthetic catalogs for multi-channel hierarchical Bayesian inference, highlighting computational aspects and issues related to hyper-prior choice. We find that when inferring stellar progenitors' properties from gravitational-wave observations, the relative efficiency in compact objects production should be taken into account. 
Finally, by simulating binary black hole detections with LIGO and Virgo sensitivity expected for the O4 observing run, we present two case studies related to the inference of the common envelope efficiency and progenitor metallicity of the binary black holes. We finally discuss how progenitors' properties can be linked to binary black hole properties.
\end{abstract}




\section{Introduction}
Since their first detection in 2015 \citep{2016PhRvL.116f1102A}, gravitational waves (GWs) have opened a new channel to study our Universe. Besides representing another confirmation of Einstein's General Relativity, GWs also provide us with a new tool for studying stellar evolution, cosmology, and the origin of compact objects.
In just 6 years from their first direct detection, and during just 3 observing runs, there has been meteoric progress in GW astrophysics. In 2017, the first Binary Neutron Star (BNS) detection with electromagnetic counterpart allowed us to measure the Hubble constant $H_0$ \citep{2017PhRvL.119p1101A, 2017Natur.551...85A}, constrain the speed of gravity, confidently link kilonovae and short $\gamma$-ray bursts and observe the formation of heavy elements via r-process  \citep{2017ApJ...848L..13A}.
The LIGO and Virgo interferometers observed GW190521, a Binary Black Hole (BBH) merger \citep{2020PhRvL.125j1102A,2020ApJ...900L..13A} with masses falling in the Pair Instability Supernova (PISN) gap. 
Another interesting example is  GW190814 \citep{2020ApJ...896L..44A}, a compact binary merger that includes a BH of $\sim 20 M_\odot$ and a secondary object falling in the expected mass gap between neutron stars and the black holes.

Interesting scientific results have also been achieved by studying the population of Compact Binaries Coalescences (CBCs). Using the GW events from the last Gravitational-Wave Transient catalogs (GWTC) \citep{2021PhRvX..11b1053A,2021arXiv211103606T}, the LIGO/Virgo/KAGRA collaboration (LVK) has been able to show that there is a smooth transition between neutron stars and black holes masses, that the preliminary BBH merger rate evolves in redshift and that the BBH mass spectrum presents several features \citep{2017ApJ...851L..25F, 2021ApJ...913L..19T,2021ApJ...913L...7A, 2021arXiv211103634T}. The LVK has been able to constrain $H_0$ using BBHs provided with galaxy catalogs \citep{2021ApJ...909..218A} and astrophysical source mass distributions \citep{2021arXiv211103604T}. All of this has been achieved with a catalog of 90 GW candidates. 

As the number of GW detections rapidly increases, population studies with GW sources are becoming a suitable tool to study the astrophysical formation channels of compact objects. Studying the population of CBCs practically consists in reconstructing the astrophysical merger rate from the observed merger rate \citep{Mandel:2018mve,2020arXiv200705579V} or vice-versa. The astrophysical merger rate is linked to astrophysical processes driving the production of the CBC population.
For instance, for BBHs, the presence of a PISN process \citep{2019ApJ...887...53F,2020ApJ...897..100V} prevents the formation of black holes (BHs) in the range $50M_{\odot}-120M_{\odot}$. See \citet{2021hgwa.bookE...4M} for an extensive review of the different formation channels for compact binaries. Population studies are also important to understand the nature of any particular ``exceptional'' event. In \citet{2010PhRvD..81h4029M, 2020PhRvD.102h3026G,2020ApJ...891L..31F, 2021PhRvD.104h3008M} the authors present a methodology to recompute the estimation of GW parameters in light of population analyses, while works such as \citet{2020ApJ...904L..26F,2021arXiv211103498F} try to reconcile GW190814 and GW190521 with the observed population of BBHs.

Currently, two methodologies are employed to reconstruct the astrophysical merger rates. The first one, which was also adopted by the LVK \citep{2020ApJ...900L..13A,2021arXiv211103634T} reconstructs merger rates in masses, spins, and redshift using inferential statistics and flexible phenomenological models. This approach is widely used in current literature \citep{2017ApJ...851L..25F,2019ApJ...886L...1V,2021arXiv211103498F,2021ApJ...922L...5C} and reconstructs the binary merger rate based on astrophysical assumptions for the phenomenological models. For instance, the BBHs merger rate as a function of redshift is usually approximated at low redshift with $(1+z)^\gamma$, in analogy to the evolution of the star formation rate at low redshift.
On one hand, this approach has the advantage of being flexible enough to fit an unknown population. On the other hand, the disadvantage of this approach is that it is not directly connected to the astrophysical processes producing BHs from their progenitors.

In order to provide a more direct astrophysical interpretation of the observed population, a parallel methodology has been employed  \citep[e.g.][]{2017ApJ...846...82Z,2017MNRAS.471.2801S,2018PhRvD..97d3014W,2019ApJ...886...25B,Delfavero2,2021MNRAS.507.5224B,2021arXiv210906222M,2021ApJ...913L...5N,2021ApJ...910..152Z,delfavero}. This methodology consists in directly reconstructing the  merger rate from astrophysical synthesis simulations of binary mergers. The central paradigm of this type of approach is to construct \textit{multi-channel} distributions, where the overall population is the sum of all the astrophysical channels simulated. As an example, one can simulate BBHs formed in isolated stellar binaries and in globular clusters and then define an overall population from them.
This type of approach has the advantage of being directly connected to the astrophysical processes forming the binary mergers but has the disadvantage of being less flexible in fitting the observed population.  

In this methodological paper, we focus on several aspects related to the interpretation and exploitation of synthetic populations of binaries for studying progenitors of GW sources.
The paper is organized as follows. In Sec.~\ref{sec:3} we provide an easy statistical method to quantify the match between phenomenological reconstructed merger rates and binary mergers catalogs. In Sec.~\ref{sec:2} we introduce key concepts for reconstructing and interpreting progenitors of mergers with multi-channel analysis. 
In Sec.~\ref{sec:4} we review and discuss critical issues of using several synthetic catalogs, that either change the astrophysical prescriptions or initial conditions, to fit observed GW events. We refer to this type of analysis as ``multi-channel reconstruction''. Differently from previous literature, we will focus on the reconstruction of stellar progenitors properties from GWs observations, showing how the relative efficiency in producing compact objects can be taken into account.
In Sec.~\ref{sec:5}, using synthetic BBHs populations, we present two case studies in which the methodologies discussed could be employed: the estimation of the common envelope efficiency and the estimation of the progenitor's metallicity. We also show how stellar progenitors' multi-channel inference can be related to the multi-channel inference of BBHs population present in litterature.
Finally, in Sec.~\ref{sec:6} we provide our final remarks. 

All results presented in this paper are generated with \textsc{gwparents}\footnote{\url{https://github.com/simone-mastrogiovanni/gwparents}}, a code for the multi-channel inference released with this work.

\section{Matching synthetic merger rates with phenomenological reconstructions}
\label{sec:3}
We first discuss in this section a quick method to quantify the agreement between synthetic binary catalogs and binary merger rates reconstructed with phenomenological models. In practice, this is the case in which we want to compare a synthetic binary catalog with a previous analysis reconstructed merger rate using phenomenological models (see \citet{2017ApJ...851L..25F,2019ApJ...886L...1V,2019PhRvD.100d3012W,2021arXiv211103498F,2021ApJ...922L...5C,2020ApJ...900L..13A,2021arXiv211103634T} as an example) from real GW events.
One qualitative avenue that was followed to perform this comparison is to ``check by eye'' the overlap of the merger rates in terms of masses and redshift of the phenomenological reconstructed rates and the synthetic catalogs. 

On one hand, this method offers a quick tool to evaluate the suitability of synthetic binaries from the phenomenological reconstruction. On the other hand, this method does not offer any statistical (or quality factor) indicator and it is hard to visualize in the case that the binary parameters are more than two.

In this section we introduce for the first time, a more quantitative method to assign a ``\textit{match}'' value to each synthetic binary model given the phenomenological reconstruction of astrophysical rates.
Let us assume that we have detected $\{x\}$ GW events from which a previous analysis estimated a posterior $p(\Lambda|\{x\})$ on some population-level parameters $\Lambda$ that describe the phenomenological rate. For instance, a population-level parameter could be the maximum mass of the BBHs mass spectrum or parameters related to the BBHs merger rate as a function of redshift.
The population-level parameters, and the phenomenological models, can be used to construct a population distribution $p_{\rm pop} (\theta|\Lambda)$, where $\theta$ represents GW source parameters such as the two masses, and a number of expected detections $N_{\rm exp}$. 
In order to assess the suitability of a synthetic population $\varphi_j$, we should compare the expected number of detections predicted by $\varphi_j$ with the one predicted from the phenomenological model. From this comparison we would like to assign a probability to each model $\varphi_j$ to fit the observed data, namely $p(\varphi_j|\{x\})$.

\begin{figure}
    \centering
    \includegraphics{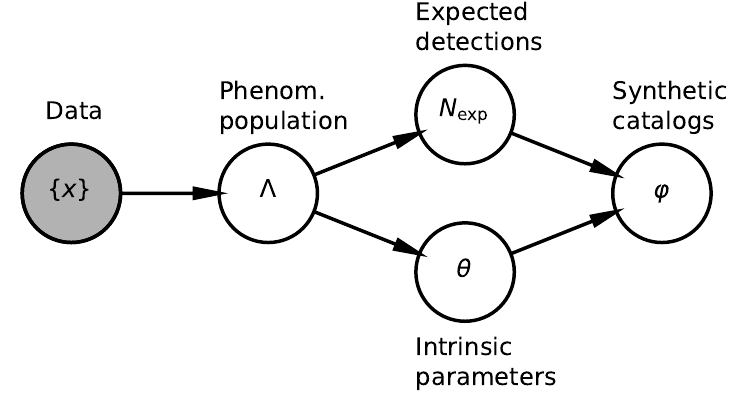}
    \caption{Bayesian graph for comparing synthetic binaries with phenomenological reconstructions of populations from real data. Each node represent an random variable, the shaded node indicates observed data. Each connection between a node and its parents indicates a conditional probability of that node given its parents.}
    \label{fig:bayesian_reco}
\end{figure}

The statistical model to compute $p(\varphi_j|\{x\})$ is depicted in Fig.~\ref{fig:bayesian_reco}.
The graph provides a quick tool for evaluating
\begin{eqnarray}
    &&p(\varphi_j|\{x\})=  \nonumber \\  &&\int p(\Lambda|\{x\})p_{\rm pop}(\theta|\Lambda)p(N_{\rm exp}|\Lambda)p(\varphi_j |\theta,N_{\rm exp}) d\Lambda dN_{\rm exp} d\theta  \nonumber \\
    &&=\int p(\Lambda|\{x\})p_{\rm pop}(\theta|\Lambda)p(\varphi_j|\theta,N_{\rm exp}(\Lambda)) d\Lambda d\theta.
    \label{eq:first}
\end{eqnarray}
In the above Eq. we have have performed the integral on $N_{\rm exp}$ by using the relation $p(N_{\rm exp}|\Lambda)=\delta(N_{\rm exp}(\Lambda)-N_{\rm exp})$, that is basically representing the fact that for each phenomenological model we can predict an expected number of GW detections. In Eq.~\eqref{eq:first}, $p(\Lambda|\{x\})$ is the posterior distribution on the phenomenological population-level parameters inferred from data and $p_{\rm pop}(\theta|\Lambda)$ the binary parameters distributions that can be reconstructed from them.
\begin{figure}
    \centering
    \includegraphics[scale=0.9]{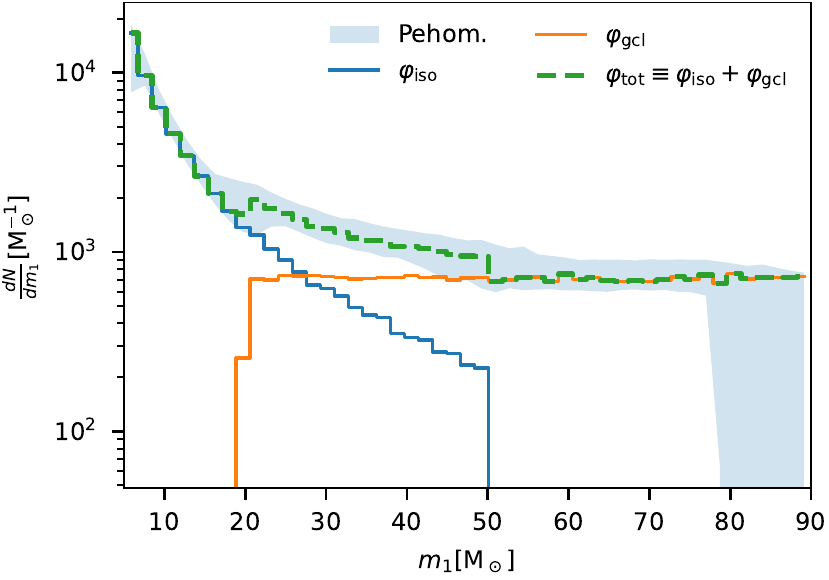}
    \caption{Primary mass distribution for the three synthetic catalogs $\varphi_{\rm iso}, \varphi_{\rm gcl}, \varphi_{\rm tot}$ and the reconstructed distribution with phenomenological model. The y-axis represent the number of binaries produced per mass bin. The figure shows qualitatively how much the synthetic merger rates overlap with the phenomenological reconstruction of the merger rate.}
    \label{fig:example}
\end{figure}
The term $p(\varphi_j|\theta,N_{\rm exp}(\Lambda))$ is a probability representing our degree of belief on the astrophysical model $\varphi_j$, given a set of binary parameters $\theta$ and expected detections $N_{\rm exp}$.
This term can be rewritten using the Bayes Theorem as 
\begin{equation}
    p(\varphi_j|\theta,N_{\rm exp}(\Lambda))= \frac{p(N_{\rm exp}(\Lambda)|\varphi_j)p_{\rm pop}(\theta|\varphi_j)p(\varphi_j)}{\sum_k p(N_{\rm exp}(\Lambda)|\varphi_k)p_{\rm pop}(\theta|\varphi_k)p(\varphi_k)},
    \label{eq:factoeye}
\end{equation}
where $p(\varphi_j)$ is a prior belief for the $j$th formation channel, $p_{\rm pop}(\theta|\varphi_j)$ is the population prior defined in Eq.~\eqref{eq:ppopsing} and  $p(N_{\rm exp}(\Lambda)|\varphi_j)$ matching the number of expected detections from the phenomenological model with the number of expected detections from the astrophysical model. 
When calculating Eq.~\eqref{eq:factoeye}, one should include also the ``complementary'' channel $\bar{\varphi}$ that covers the parameter space $\theta$ not covered by any of the other channels, i.e. $p(\bar{\varphi}|\theta,N_{\rm exp}(\Lambda))=1-\sum_j p(\varphi_j|\theta,N_{\rm exp}(\Lambda))$. Note  that Eq.~\eqref{eq:factoeye} reduces to the ratio of the population priors in the limit that all the models predict the same number of expected detections. Note also that in this analysis we do not need to include selection biases as they have already been deconvolved by the analysis that fit the phenomenological model. In other words, we are comparing astrophysical rates and not observed rates.

Eq.~\eqref{eq:first} can be computed using the following procedure: if we are provided with a set of $N_{\Lambda}$ posterior samples for the population phenomenological parameters $\Lambda_i$, for each $\Lambda_i$ one  can compute the expected number of events $N_{\rm exp}(\Lambda_i)$, then draw $N_{\theta}$ binaries from the population distribution $p_{\rm pop}(\theta|\Lambda_i)$ and evaluate the integral in Eq.~\eqref{eq:first} as

\begin{equation}
    p(\varphi_j|\{x\})=\frac{1}{N_{\Lambda}N_{\theta}}\sum_{i}^{N_{\Lambda}}\sum_{k}^{N_{\theta}}p(\varphi_j|\theta_k,N_{\rm exp}(\Lambda_i)).
    \label{eq:sum1}
\end{equation}

Let us give an example. We simulate two populations of BBHs that we refer to ``isolated'' ($\varphi_{\rm iso}$) and ``globular clusters'' ($\varphi_{\rm gcl}$) in analogy with the current BBHs formation channels reviewed in \citep{2021hgwa.bookE...4M}. 
The $\varphi_{\rm iso}$ population produces a total of $10^5$ BBHs with primary mass $m_1$ distributed according to a truncated power law $p(m_1) \propto m_{1}^{-2}$ between $5 M_\odot$ and $50 M_\odot$, while $m_2$ is distributed between $5 M_{\odot}$ and $m_1$ with a power law $p(m_2|m_1) \propto m_2$. The $\varphi_{\rm gcl}$ produces a total of $5 \cdot 10^3$ BBHs with primary mass uniform in $20 M_{\odot}$ and $90 M_{\odot}$ and secondary mass uniform in  $5 M_{\odot}$ and $m_1$. The overall population of BBHs is defined as the sum of the two channels, i.e $\varphi_{\rm tot}=\varphi_{\rm iso}+\varphi_{\rm gcl}$. 
We also assume that a previous analysis using BBHs from the $\varphi_{\rm tot}$ population has been able to fit the mass spectrum with a broken power and obtained a $10\%$ error on the mass spectrum parameters and overall merger rate. 
The three populations $\varphi_{\rm iso}, \varphi_{\rm gcl}, \varphi_{\rm tot}$ and the phenomenological reconstruction of $\varphi_{\rm tot}$ are represented in Fig.~\ref{fig:example}. 
The figure shows how $ \varphi_{\rm tot}$ overlaps with the phenomenological reconstruction. While $\varphi_{\rm iso}, \varphi_{\rm gcl}$ fit only the total population in the low and high mass regions with an overlap between $20 M_{\odot}$ and $50 M_{\odot}$. We now want to assess the three models $\varphi_{\rm iso}, \varphi_{\rm gcl}, \varphi_{\rm tot}$ with the reconstructed population and find which one is preferred.

\begin{figure*}
    \centering
    \begin{subfigure}[b]{0.475\textwidth}
        \centering
        \includegraphics[width=\textwidth]{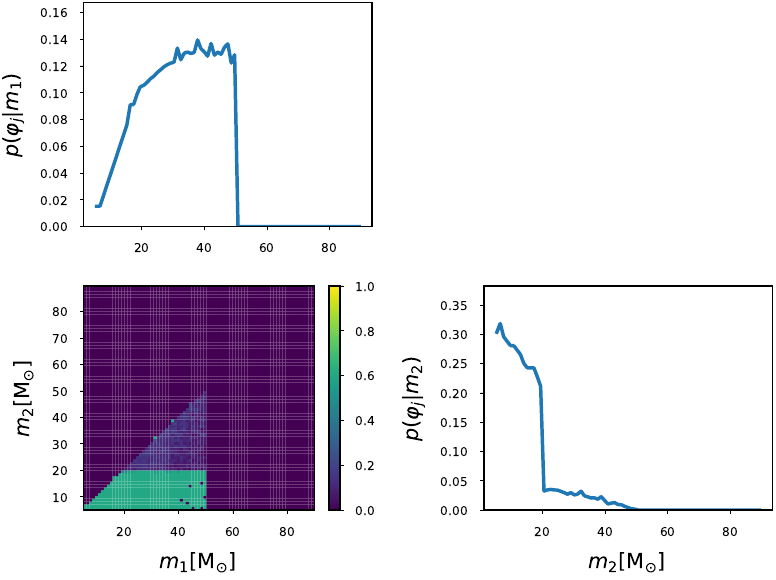}
        \caption[Network2]%
        {{\small Model $\varphi_{\rm iso}$}}    
        \label{fig:mean and std of net14}
    \end{subfigure}
    \hfill
    \begin{subfigure}[b]{0.475\textwidth}  
        \centering 
        \includegraphics[width=\textwidth]{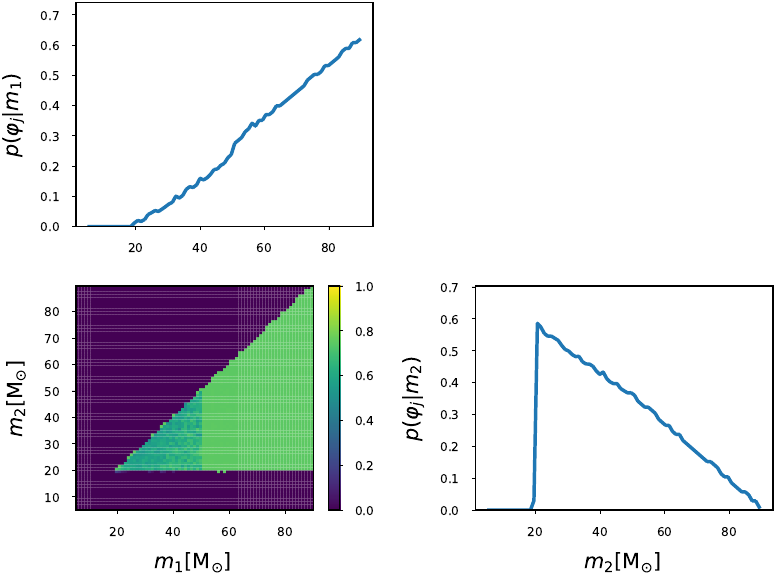}
        \caption[]%
        {{\small Model $\varphi_{\rm gcl}$}}    
        \label{fig:mean and std of net24}
    \end{subfigure}
    \vskip\baselineskip
    \begin{subfigure}[b]{0.475\textwidth}   
        \centering 
        \includegraphics[width=\textwidth]{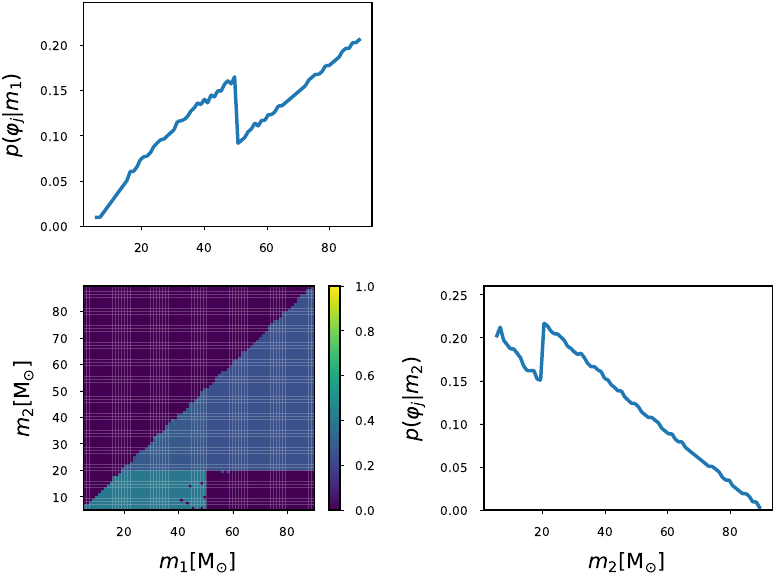}
        \caption[]%
        {{\small Model $\varphi_{\rm tot}$}}    
        \label{fig:mean and std of net34}
    \end{subfigure}
    \hfill
    \begin{subfigure}[b]{0.475\textwidth}   
        \centering 
        \includegraphics[width=\textwidth]{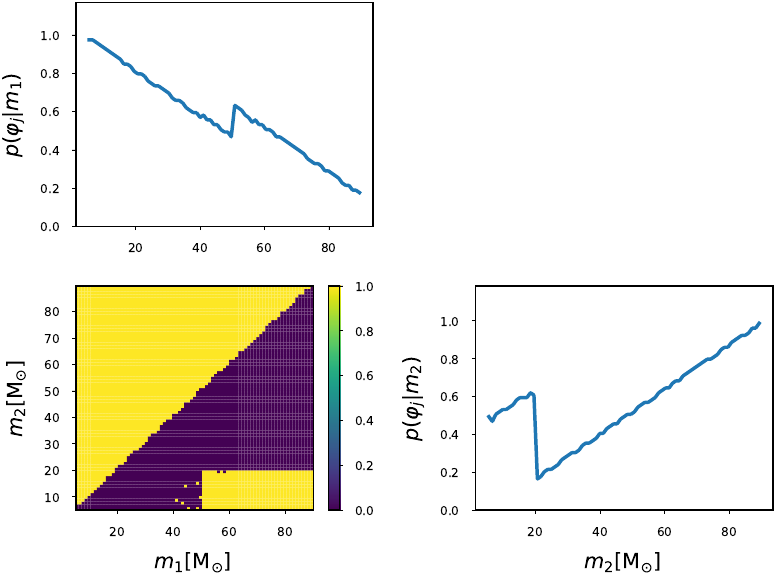}
        \caption[]%
        {{\small Model $\bar{\varphi}$}}    
        \label{fig:mean and std of net44}
    \end{subfigure}
    \caption[ The average and standard deviation of critical parameters ]
    {\small Plots of $p(\varphi_j|\theta,N_{\rm exp}(\Lambda))$ given in Eq.~\eqref{eq:factoeye}. for the three synthetic catalogs as a function of the masses. The fourth model is the complementary model and represent the complementary model in the masses space to all the other models. The figures have been generated assuming that all the models predicts the same number of BBHs. This choice was made as in this limit Eq.~\eqref{eq:factoeye} reduces to the ratio of the population probabilities of each model. The figures have been generated by dividing the mass space in 6400 bins equally sized. The colorbars indicates the values of models probability given a mass value.} 
    \label{fig:bione}
\end{figure*}

The first ingredient that we need, is the evaluation of Eq.~\eqref{eq:factoeye} as a function of the BBH masses. In Fig.~\ref{fig:bione} we show $p(\varphi_j|m_1,m_2)$ computed for all the formation channels.
The figure shows the interpretation of $p(\varphi_j|m_1,m_2)$: when we have mass values in the range $m_{1,2}<20 M_{\odot}$, the most probable formation channel is $\varphi_{\rm iso}$, while when we are looking at binaries with $m_{1,2}>50$ the most probable formation channel is $\varphi_{\rm gcl}$.
It is also interesting to note that the complementary formation channel  is  $100\%$ probable  where none of the models considered produces masses, e.g. for the region $m_2>m_1$ which is excluded by our simulation. With an evaluation of $p(\varphi_j|m_1,m_2)$, we can now calculate Eq.~\eqref{eq:sum1} by using samples from the phenomenological reconstructed rate. For didactic purposes, let us consider two cases. In the first, we will assume that each formation channel predicts the same amount of BBHs; the preference is solely given by comparing the different mass distributions. In the second, we will include information on how many BBHs each formation channel predicts.

In the first case we obtain $p(\varphi_{\rm iso}|\{x\})=34\%, p(\varphi_{\rm gcl}|\{x\})=25\%, p(\varphi_{\rm tot}|\{x\})=34\%, p(\bar{\varphi}|\{x\})=7\%$. These probabilities can be used to evaluate how much the population probabilities $p(\theta|\varphi_j)$ overlap with the population probability of the phenomenological rates $p(\theta|\Lambda)$. 
For instance, $\varphi_{\rm tot}$  fits 1.36 times better the distribution of masses with respect to $\varphi_{\rm gcl}$.
If we now include the fact that each formation channel predicts a different amount of BBHs produced, we obtain $p(\varphi_{\rm iso}|\{x\})=0.5\%, p(\varphi_{\rm gcl}|\{x\})=0.5\%, p(\varphi_{\rm tot}|\{x\})=91\%, p(\bar{\varphi}|\{x\})=8\%$. The clear preference for the $\varphi_{\rm tot}$ channel is now given by the fact that the number of BBHs produced by $\varphi_{\rm iso}$ and $\varphi_{\rm gcl}$ alone is not enough alone to match the total number of BBHs reconstructed by the phenomenological model.

So far, we have discussed a quantitative method to compare synthetic binary catalogs with phenomenological merger rate reconstructions. This method evaluates the overlap of each model by considering it independent from the others. In the next sections, we will focus on analyses that aim at reconstructing the binary merger rates as a combination of the progenitors' simulation at our disposal.

\section{Building multi-channel merger rates from the black holes progenitors}
\label{sec:2}
In this section, we follow a top-to-bottom derivation to show how it is possible to build binary merger rates from synthetic binary catalogs. In the rest of this work, we will focus mostly on BBHs. 

Let us assume that we have generated a population of BBHs progenitors $N_{*}^{\varphi_j}$ that we evolve through an astrophysical channel $\varphi_j$ to obtain a certain number of BBH mergers $N_{\rm BBH}^{\varphi_j}$.
The merger rate of BBHs for each astrophysical channel can be written as 
\begin{equation}
    \frac{d N^{\varphi_{j}}_{\rm BBH}}{ d\theta dz dt}=\mathcal{T}^{\varphi_{j}}(\theta,\theta_*,z,z_*,t,t_*)\frac{d N^{\varphi_{j}}_{*}}{ d\theta_* dz_* dt_*},
    \label{eq:prog}
\end{equation}
where  $z_*$ is the redshift at which the BBH progenitor is formed, $\theta_*$ a set of the progenitor parameters such as metallicity and $dt_*$ indicates the time interval at the progenitor redshift. The function $\mathcal{T}$ can be understood as a ``transfer function'' that tells us if a progenitor with parameters $\theta_*$ at redshift $z_*$ would produce a BBH with parameters $\theta$ at redshift $z$. A central quantity for many population analyses is the \textit{population probability}  that is built from the binary merger rate as
\begin{equation}
 p_{\rm pop}(\theta,z,t|\varphi_j)= \frac{1}{N^{\varphi_j}_{\rm BBH}} \frac{d N^{\varphi_{j}}_{\rm BBH}}{ d\theta dz dt},
 \label{eq:ppopsing}
\end{equation}
where the term $N^{\varphi_j}_{\rm BBH}$ is the total number of BBHs predicted by the formation channel $\varphi_j$. 

The idea behind multi-channel analysis \citep{2017MNRAS.471.2801S,2017ApJ...846...82Z,2018PhRvD..97d3014W} is to construct (and compare with observed events) an overall BBH merger rate, built as a linear combination of various formation channels, namely
\begin{equation}
    \frac{d N_{\rm BBH}}{ d\theta dz dt} = \sum^{N_{\rm syn}}_{j} \lambda_j \frac{d N^{\varphi_{j}}_{\rm BBH}}{ d\theta dz dt}.
    \label{eq:overrate}
\end{equation}
The $\lambda_j$ coefficients are a set of \textit{mixture coefficients}, that are usually fit in the analysis. The rationale behind this idea is that one single formation channel could not be sufficient to describe the population of observed BBHs (e.g. in the case that BBHs are formed from isolated binary evolution or in globular clusters).

Let us now comment on the physical interpretation for the $\{\lambda\}$ coefficients and their relation to the construction of synthetic binary catalogs. These terms can be understood in terms of progenitors' population.
Using Eqs.~\eqref{eq:prog}-\eqref{eq:overrate}, the overall BBHs merger rate can be written as
\begin{equation}
    \frac{d N_{\rm BBH}}{ d\theta dz dt_s} = \sum_j  \mathcal{T}^{\varphi_{j}}(\theta,\theta_*,z,z_*,t,t_*) \lambda_j\frac{d N^{\varphi_{j}}_{*}}{ d\theta_* dz_* dt_*}.
    \label{eq:tfun}
\end{equation}
If we assume that the BBHs progenitors distribution is in common across all the formation channels considered, then the set of $\{\lambda\}$ should respect the condition $\sum_j \lambda_j=1$. Namely, the $\{\lambda\}$ represents the fraction of progenitors that produce BBHs through the formation channels $\{\varphi\}$. In this case we refer to the $\{\lambda\}$ as \textit{fractional mixture coefficients}.


A completely different case can be found when we want to fit the observed BBHs using synthetic catalogs generated from \textit{independent} populations of progenitors. Therefore, the overall population of BBHs will be given by the sum of all the independent sub-populations. 
In this case,  the set $\{\lambda\}$ represents the abundance of each subpopulation of BBHs in the observed data. If the observed data is correctly described by the modeled BBHs sub-populations, we would expect each $\lambda_j=1$. Values of $\lambda_j>1$ will either indicate that the sub-populations are more numerous or that the transfer function is twice more effective in producing BBHs from the progenitors. The opposite is true for values of $\lambda_j<1$.

From Eq.~\eqref{eq:overrate} it is possible to define a population probability given as
\begin{equation}
 p_{\rm pop}(\theta,z,t|\{\lambda\varphi\})= \frac{1}{N_{\rm BBH}} \frac{d N_{\rm BBH}}{ d\theta dz dt},
\end{equation}
where with $\{\lambda \varphi \}$ we indicate a collection of formation channels multiplied by their mixture coefficients.
By using Eq.~\eqref{eq:overrate} and the fact that $N_{\rm BBH}=\sum_j \lambda_j N^{\varphi_{j}}_{\rm BBH}$, one can show that the overall population probability is
\begin{equation}
    p_{\rm pop}(\theta,z,t|\{\lambda\varphi\})= \sum_j  \frac{\lambda_j N^{\varphi_{j}}_{\rm BBH}}{\sum_k \lambda_k N^{\varphi_{k}}_{\rm BBH}} p_{\rm pop}(\theta,z,t|\varphi_j).
    \label{eq:res1}
\end{equation}
The  equation above has a direct astrophysical interpretation: if we are provided with a formation channel $\varphi_j$ that predicts significantly more BBHs than the others, then the overall population probability must be dominated by this channel. 
The term 
\begin{equation}
p(\varphi_j|\{\lambda\varphi\})=\frac{\lambda_j N^{\varphi_{j}}_{\rm BBH}}{\sum_k \lambda_k N^{\varphi_{k}}_{\rm BBH}}   
\label{eq:missing}
\end{equation}
can be also understood as a probability of the model $\varphi_j$ given the scalar coefficients $\{\lambda\}$ and the other models $\{\varphi\}$. With this definition, Eq.~\eqref{eq:res1} can be written as
\begin{equation}
    p_{\rm pop}(\theta,z,t|\{\lambda\varphi\})= \sum_j  p(\varphi_j|\{\lambda\varphi\}) p_{\rm pop}(\theta,z,t|\varphi_j).
    \label{eq:res2}
\end{equation}

\begin{figure}
    \centering
    \includegraphics[scale=0.5]{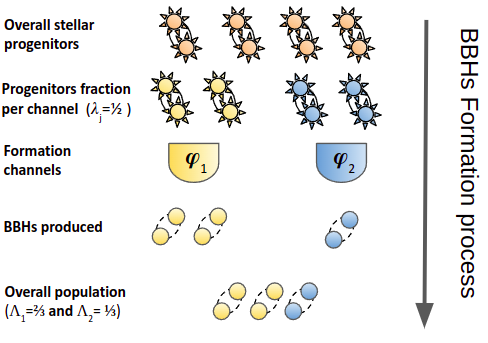}
    \caption{The illustration shows the relation between the coefficients $\lambda_j$ and $\Lambda_j$ by following the BBHs from the stellar progenitors. The figure starts from a common population of stellar progenitors, half of which enter a formation channel $\varphi_1$ (yellow) and the other half $\varphi_2$ (blue). The first formation channel is two times more efficient than the second in producing BBHs. At the end $2/3$ of the observable population of BBHs has been produced in $\varphi_1$ and 1/3 in $\varphi_2$.}
    \label{fig:scheme}
\end{figure}

Note that there is a fundamental difference between the construction of the above population probability and the one used in several recent works such as \citet{2017MNRAS.471.2801S,2017ApJ...846...82Z,2019ApJ...886...25B,2021arXiv210906222M,2021MNRAS.507.5224B,2021PhRvD.103h3021W}. In these works, the multi-channel population probability is built as 
\begin{equation}
    p_{\rm pop}(\theta,z,t|\{\Lambda\})= \sum_j  \Lambda_j p_{\rm pop}(\theta,z,t|\varphi_j),
    \label{eq:wrong}
\end{equation}
where $\sum \Lambda_j=1$. The parameters $\Lambda_j$ effectively represent the fraction of the BBH distribution given by a particular formation channel. Instead, the $\lambda_j$ defined in this paper represent the fraction of progenitors producing BBHs in a given formation channel. In order to define a progenitor-induced BBH population probability, it is important to take into account the term $p(\varphi_j|\{\lambda\varphi\})$.
In Sec.~\ref{sec:bias} we provide an example to discuss how these two quantities are related and can be converted to each other. For now let us give a simple example illustrated in Fig.~\ref{fig:scheme} to better understand the meaning of the $\lambda_j$ and $\Lambda_j$ coefficients.

Let us assume to be provided with a set of progenitors producing BBHs via two formation channels $\varphi_1$ and $\varphi_2$, with the first formation channel predicting 2 times more BBHs than the other. Let us also assume that 1/2 of progenitors enter the first formation channel and 1/2 of the second. In other words $\lambda_1=\lambda_2=1/2$. When we look at the population distribution of BBHs, we would find that 2/3 of the BBHs are produced in the formation channel $\varphi_1$ and 1/3 by $\varphi_2$. In other words $\Lambda_1=2/3, \Lambda_2=1/3$. Therefore, if we perform our inference using Eq.~\eqref{eq:wrong}, we can not directly use the $\Lambda_j$ to draw conclusions about the BBHs progenitors. We can only draw conclusions about the fraction of BBHs produced in a given formation mechanism.

\section{Progenitors multi-channel Bayesian analyses}
\label{sec:4}

We now discuss the case in which we would like to reconstruct the merger rate for multiple formation channels starting from the observed BBHs. Differently from what we discussed in the previous section, in this case, we will not use phenomenological models and we will rely solely on synthetic binary catalogs. 
We will use the mixture model approach presented in Sec.~\ref{sec:2} and write the overall BBHs merger rate as in Eq.~\eqref{eq:overrate}. 
We will discuss in Sec.~\ref{sec:4.1} the statistical background for multi-channel analyses based on synthetic catalogs, in Sec.~\ref{sec:4.2} how priors on  the mixture coefficients can be chosen and in Sec.~\ref{sec:4.3} computational difficulties related to this kind of analysis.

\begin{figure*}
    \centering
    \begin{subfigure}[b]{0.24\textwidth}
        \centering
        \includegraphics[width=\textwidth]{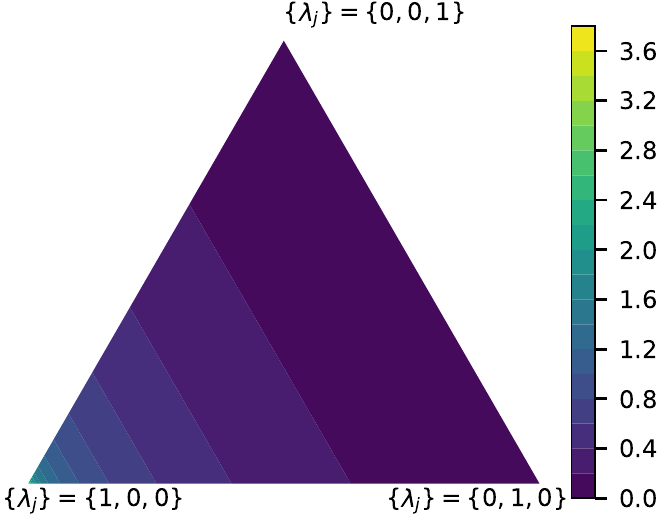}
        \caption[Network2]%
        {{\small Conditional Uniform prior}}    
        \label{fig:mean and std of net14}
    \end{subfigure}
    \begin{subfigure}[b]{0.24\textwidth}  
        \centering 
        \includegraphics[width=\textwidth]{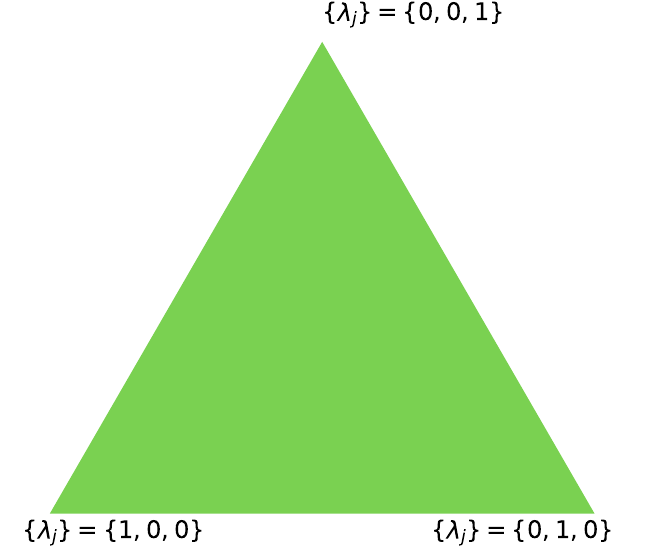}
        \caption[]%
        {{\small ``Flat'' Dirichlet}}    
        \label{fig:mean and std of net24}
    \end{subfigure}
    \begin{subfigure}[b]{0.24\textwidth}   
        \centering 
        \includegraphics[width=\textwidth]{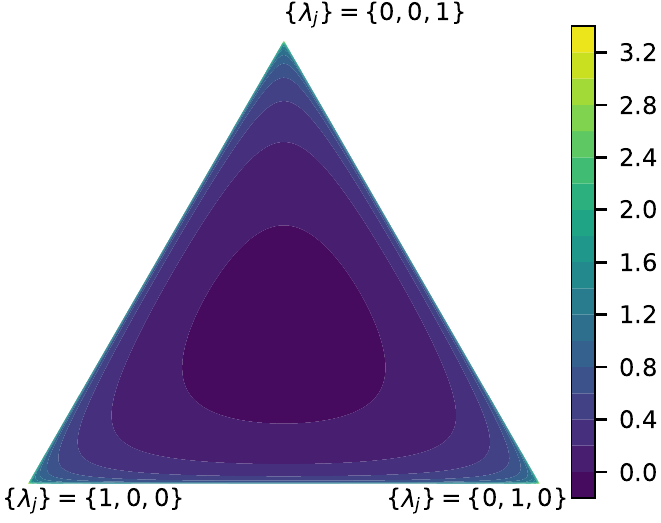}
        \caption[]%
        {{\small Dirichlet favoring single models}}    
        \label{fig:mean and std of net34}
    \end{subfigure}
    \begin{subfigure}[b]{0.24\textwidth}   
        \centering 
        \includegraphics[width=\textwidth]{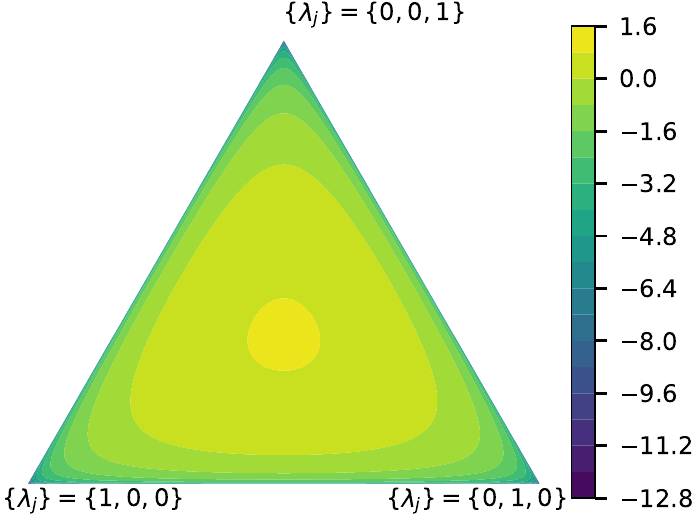}
        \caption[]%
        {{\small Dirichlet favoring mixture models}}    
        \label{fig:mean and std of net44}
    \end{subfigure}
    \caption[ The average and standard deviation of critical parameters ]
    {\small Representation on a 2-simplex of the logarithm of the prior probability distribution on the fractional mixture coefficients $\{\lambda\}$ of three models. From left two right: Conditional uniform prior with ordering preference, ``Flat'' Dirichlet prior with concentration parameters $1$, Dirichlet prior favoring single models with concentration parameters $0.5$ and Dirichlet prior favoring mixture models with concentration parameters $3.0$.} 
    \label{fig:ciao}
\end{figure*}

\subsection{Statistical method}
\label{sec:4.1}

The hierarchical likelihood of having $N_{\rm obs}$ GW events from data $\{x\}$ conditioned on the set of models $\varphi_j$ and the mixture coefficients $\lambda_j$ is (see \citet{2020arXiv200705579V} for a bottom-to-top derivation)
\begin{equation}
    p(\{x\}|\{\lambda\varphi\}) \propto e^{-N_{\rm exp}} \prod_i^{N_{\rm obs}} T_{\rm obs} \int  \frac{ p(x_i|\theta)}{1+z} \frac{dN_{\rm BBH}}{d\theta dz dt}  d\theta dz,
    \label{eq:pos}
\end{equation}
where $p(x_i|\theta)$ is the GW likelihood and $N_{\rm exp}$ is the number of expected events observable in a given observing time $T_{\rm obs}$. The GW likelihood  quantifies the uncertainties with which the source astrophysical parameters, such as luminosity distance and detector-frame masses, are determined.
Eq.~\eqref{eq:pos} can be rewritten in the alternative form \citep{2020arXiv200705579V}
\begin{equation}
    p(\{x\}|\{\lambda\varphi\}) \propto e^{-N_{\rm exp}} N_{\rm exp}^{N_{\rm obs}} \prod_i^{N_{\rm obs}} \frac{\int p(x_i|\theta) p_{\rm pop}(\theta|\{\lambda\varphi\}) d\theta}{\beta(\{\lambda\varphi\})},
    \label{eq:pos1}
\end{equation}
where  $p_{\rm pop}(\theta|\{\lambda\varphi\})$ is the population probability defined as in Eq.~\eqref{eq:res2}, and $\beta(\{\lambda\varphi\})$ is the selection effect (see later).

Our aim is to quickly evaluate Eq.~\eqref{eq:pos1} as a function of the mixture coefficients $\lambda_j$. We will factorize Eq.~\eqref{eq:pos1} in several terms that can be computed once for each formation channel $\varphi_j$ and rescaled with $\lambda_j$ to quickly evaluate the hierarchical likelihood.  The numerator factor in the product of Eq.~\eqref{eq:pos1} can be rewritten as,
\begin{equation}
    \int p(x_i|\theta) p_{\rm pop}(\theta|\{\lambda\varphi\}) d\theta=\sum_j p(\varphi_j|\{\lambda\varphi\}) \mathcal{L}_{i,j},
    \label{eq:ciao}
\end{equation}
where we have expanded $p_{\rm pop}(\theta|\{\lambda\varphi\})$ using Eq.~\eqref{eq:res2} and we have defined 
\begin{eqnarray}
    \mathcal{L}_{i,j}=\int p(x_i|\theta) p_{\rm pop}(\theta|\varphi_j) d\theta.
    \label{eq:matchl}
\end{eqnarray}
The $\mathcal{L}_{i,j}$ can be evaluated numerically once for each $i{\rm th}$ GW event and $j{\rm th}$ formation channel. We also recall that $p(\varphi_j|\{\lambda\varphi\})$ can be constructed using Eq.~\eqref{eq:missing} and using only the number of BBHs predicted by each model and the mixture coefficients $\lambda_j$. 
The selection effect $\beta(\{\lambda\varphi\})$ can be quickly computed by knowing the total number of BBHs predicted by each formation channel and the fraction of BBHs detectable by each channel $\beta(\varphi_j)$, namely
\begin{equation}
    \beta(\{\lambda\varphi\})=\frac{\sum_j \lambda_j N_{\rm BBH}^{\varphi_j} \beta(\varphi_j)}{\sum_j \lambda_j N_{\rm BBH}^{\varphi_j}}.
    \label{eq:sel}
\end{equation}

Finally, the Poissonian term $$e^{-N_{\rm exp}} N_{\rm exp}^{N_{\rm obs}}$$ in Eq.~\eqref{eq:pos1} can be easily computed by recognizing that $N_{\rm exp}=\sum_j \lambda_j N_{\rm BBH}^{\varphi_j} \beta(\varphi_j)$. 

This term is usually marginalized out  in multi-channel analyses focusing on BBHs population as performed in \citet{2017ApJ...846...82Z,2017MNRAS.471.2801S,2021arXiv210906222M,2019ApJ...886...25B,2021MNRAS.507.5224B,2021ApJ...910..152Z}. To do so, we need to introduce a ``nuisance scaling parameter'' $A$ in common to all the population models such that $N_{\rm BBH}= A \sum_j \lambda_j N_{\rm BBH}^{\varphi_j}$. If we take a prior on $A$ uniformly distributed in logarithmic space, it is possible to marginalize out the Poissonian term and Eq.~\eqref{eq:pos1} reduces to
\begin{equation}
    p(\{x\}|\{\lambda\varphi\}) \propto  \prod_i^{N_{\rm obs}} \frac{\int p(x_i|\theta) p_{\rm pop}(\theta|\{\lambda\varphi\}) d\theta}{\beta(\{\lambda\varphi\})}.
    \label{eq:pos3}
\end{equation}

Note that, in comparison to population analyses based on phenomenological models \citep{2017ApJ...851L..25F,2019ApJ...886L...1V,2021arXiv211103498F,2021ApJ...922L...5C,2020ApJ...900L..13A,2021arXiv211103634T}, the parameter $A$ effectively act as a common rescaling for the BBH merger rate density $R_0(\varphi_j)$ identified by each model. In other words, Eq.~\eqref{eq:pos3} reconstructs the BBHs astrophysical distributions in terms of masses and redshift without accounting for the absolute merger rate. While this choice is mathematically correct and reconstructs the correct distribution in masses and redshift of BBHs, one should be careful about the astrophysical interpretation. For instance, the synthetic simulations might predict many more events than the observed ones, while still being able to fit the redshift and mass distribution. This choice is usually done when the rates of the different formation channels are highly uncertain, but the mass and redshift distributions are not.

To summarize, in order to quickly perform a multi-channel analysis using several formation channels $\varphi_j$, we  need to: \textit{(i)} Estimate the total number of BBHs produced by each formation channel $N_{\rm BBH}^{\varphi_j}$ and their detectable fraction $\beta(\varphi_j)$, \textit{(ii)} for each formation channel and GW event estimate the term $\mathcal{L}_{i,j}$ in Eq.~\eqref{eq:matchl} and \textit{(iii)} for some values of the set $\{\lambda\}$ use Eq.~\eqref{eq:ciao} and Eq.~\eqref{eq:sel} to effectively build the hierarchical likelihood.

\subsection{Priors on the mixture coefficients}
\label{sec:4.2}

We now discuss how priors on the mixture coefficients $\{\lambda\}$ can be chosen according to the astrophysical case considered.
In Sec.~\ref{sec:2} we considered  two cases: the case in which each formation channel has an independent sub-population of progenitors and the case for which the population of BBHs progenitors is in common to each formation channel. In the former,  the $\lambda_j$ are independent of each other and a value of $\lambda_j=1$ indicates that the BBH formation channel is observed in data as the model predicts. In this case, each prior on $\lambda_j$ can be chosen independently and from an astrophysical point of view, this case corresponds to changing the initial astrophysical conditions (e.g star formation rate) of the simulation.
In the latter,  the set of $\{\lambda\}$ must satisfy the constraint $\sum_j \lambda_j=1$, and these parameters effectively represent the relative fraction of BBHs observed produced by each channel. In terms of astrophysics, this represents the case for which the initial conditions of the simulation are set but the astrophysical evolution prescriptions are changed.

\begin{figure}
    \centering
    \includegraphics{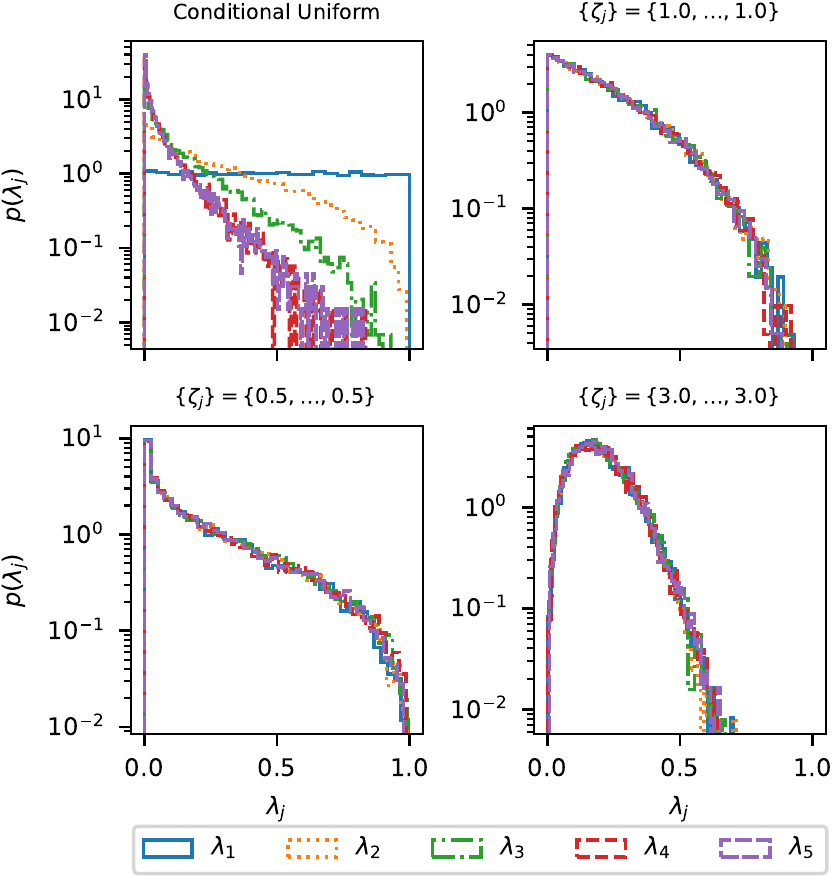}
    \caption{Marginal priors on $5$ mixture models. The different panels correspond to the marignal priors for different choices as described in Fig.~\ref{fig:ciao} and Sec.~\ref{sec:4.1}. The conditional uniform prior is given in Eq.~\eqref{eq:mess}.}
    \label{fig:prior_effect}
\end{figure}

When the $\{\lambda\}$ must be normalized (common BBHs progenitors), a non-trivial bound is introduced in the joint prior of the mixture coefficients. One possibility is to build a joint prior that satisfies the normalization constraint by drawing sequentially the values of $\lambda_j$ from a cascade of conditional probabilities. Namely, we write the joint prior as 
\begin{equation}
    p(\{\lambda\})=\prod_j p(\lambda_j|\{\lambda\}_{i<j}),
    \label{eq:mess}
\end{equation}
where $\{\lambda\}_{i<j}$ indicates a set of $\lambda_i$ with index lower than $j$.
By choosing uniform conditional priors, the above equation can be written as
\begin{eqnarray*}
&&p(\lambda_1)=p(\lambda_1) \\
&&p(\lambda_j|\{\lambda\}_{i<j})=
  \begin{cases}
   p(\lambda_j|\{\lambda\}_{i<j}) &  \text{if } 0 \leq \lambda_j \leq 1-\sum_{i<j} \lambda_i \\
    0 & \text{if }   \lambda_j > 1-\sum_{i<j} \lambda_i
  \end{cases} \\
&&p(\lambda_N|\{\lambda\}_{i<N})=\delta(1-\sum_{i<N} \lambda_i). 
\end{eqnarray*}
We note that this prior choice is not optimal for multi-channel studies as it introduces an ordering preference. In Fig.~\ref{fig:ciao} (left panel), we show the logarithm of a joint prior to built in this way for the case that we are provided with three astrophysical formation channels. As it can be seen from the figure, this  prior naturally prefers the first ordered model. In the case that multiple models are provided, this type of prior will strongly disfavor models that are ordered in the last. We display this effect in Fig.~\ref{fig:prior_effect} by showing the marginal prior distributions in the case that we are provided with 5 astrophysical channels. 

A more natural choice that removes the problem of model ordering is to use a Dirichlet distribution on the $\{\lambda\}$ as done in \citet{2017MNRAS.471.2801S,2017ApJ...846...82Z,2018PhRvD..97d3014W}. The Dirichlet distribution ensures the normalization of the $\{\lambda\}$ and also provides a set of concentration parameters $\{\zeta\}$ governing how the probability is distributed on the plane identified by $\sum \lambda_j=1$. Fig.~\ref{fig:ciao} shows the logarithm of the Dirichlet prior to different choices of the $\{\zeta\}$ parameters. If $\zeta_j=1$, the prior probability is uniform across the combination of all the formation channels. If $\zeta_j<1$, the prior will prefer to build the overall BBH rate using a single formation model. Finally, if $\zeta_j>1$, the prior will prefer to build the BBH rate as a superposition of all the models. In general, as we will show in Sec.~\ref{sec:5.1}, the $\{\zeta\}$ parameters can also be treated as free parameters to infer. The marginal priors on the $\lambda_j$ in the cases presented for a Dirichlet distribution are shown in Fig.~\ref{fig:prior_effect}. One can observe that the marginal priors are equal for all the models. 

We therefore argue that Dirichlet priors should be used when performing this type of analyses.

\subsection{Evaluating Monte Carlo integrals}
\label{sec:4.3}
    
The calculation of the hierarchical likelihood in Eq.~\eqref{eq:pos} requires the evaluation of several numerical integrals. A first implicit integral is given by the calculation of the fraction of BBHs that we expect to detect. This integral is given by the product of the detection probability as a function of the BBHs parameters with the BBHs population distribution and merger rate. The integral is not evaluated analytically and a common technique to estimate it is by using injection studies. The idea is simply to generate GW injections in noise from the desired BBH population and estimate what is the detectable fraction.  \citet{2019RNAAS...3...66F} showed that the number of detectable injections should be at least 4 times higher than the GW events considered in the analysis. Otherwise, the evaluation of the hierarchical likelihood is not numerically stable.

The other term that requires an integral over the BBH population is Eq.~\eqref{eq:matchl}. This integral it is usually evaluated as a \textit{Monte Carlo integral} by using $N_s$ samples from the posterior of each BBH detected.
This approach consists in approximating the integral as
\begin{equation}
    \mathcal{L}_{i,j} \approx \frac{1}{N_s} \sum_j^{N_s} \frac{p_{\rm pop}(\theta|\varphi_j)}{p_0(\theta)},
    \label{eq:int1}
\end{equation}
where $p_0(\theta)$ is a prior applied to calculate the BBH posteriors on the binary parameters and $p_{\rm pop}(\theta|\varphi_j)$ is the population prior associated to the formation channel. Alternatively, one can decide to perform the Monte Carlo integral by summing over simulated BBHs from the formation channel $p_{\rm pop}(\theta|\varphi_j)$ and write
\begin{equation}
    \mathcal{L}_{i,j} \approx \frac{1}{N_s} \sum_j^{N_s} \frac{p(\theta|x_i)}{p_0(\theta)}.
    \label{eq:int2}
\end{equation}
In principle, we would expect Eq.~\eqref{eq:int1} and Eq.~\eqref{eq:int2} to return the same result. 
Both approaches have in common one necessity, either the BBH population of the formation channel or the GW posterior of observed events should be known as a function of the parameters $\theta$. These analytic functions are not usually known, in fact, we are usually provided with either a list of posterior samples from $p(\theta|x_i)$ or a list of BBHs simulated from $p_{\rm pop}(\theta|\varphi_j)$. One possibility to compute analytically this probability from a set of samples, is by using kernel density estimates, histograms, or non-parametric fitting such as the ones proposed in \citet{2018PhRvD..97d3014W,2022ApJ...926...79G,2018MNRAS.479..601D, 2018ApJ...868..140T, 2021arXiv211212659S, Delfavero2, 2022MNRAS.509.5454R, delfavero}. If  Eq.~\eqref{eq:int1} is used, the sum is performed over GW posterior samples and the fit is on the BBH population of the formation channel. If Eq.~\eqref{eq:int2} is used, the sum is over the BBHs predicted by the formation channel and the GW posterior is evaluated by the fit. In both cases, a possible gaussian kernel fitting should always be  validated against the original distribution. 

A rule of thumb  to decide what is best suited to evaluate the Monte Carlo Integral is the following. Eq.~\eqref{eq:int1} can be used if the formation channel has a phenomenological (or semi-analytical) model and no fitting is needed.  Eq.~\eqref{eq:int1} can also be used when the BBH parameters from GW data are measured with a precision significantly lower than the typical ranges covered by the BBH formation channels.
Eq.~\eqref{eq:int2} can be used in this case that the range spanned by the BBH formation channels is comparable or significantly lower than the precision with which we are able to measure BBH parameters from data.

Current population studies are based on BBHs, for which we expect formation channels to cover a wider range in masses and redshift to the typical error budgets estimated from GW data. That is why so far Monte Carlo Integrals are mostly evaluated with Eq.~\eqref{eq:int1}.

\begin{figure*}
    \centering
    \includegraphics{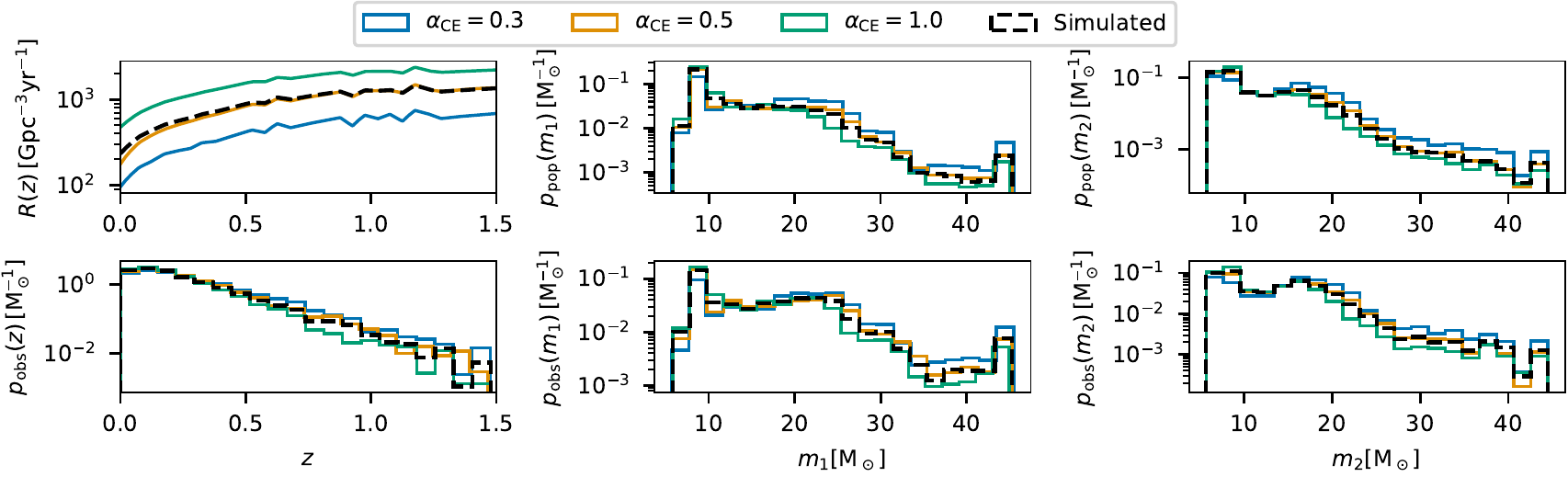}
    \caption{\textit{Left:} BBH merger rate redshift evolution for the three common envelope models we consider. \textit{Center:} Distribution of the primary source mass for the models we consider. \textit{Right:} Distribution of the secondary source mass for the models we consider. In all the panels, the simulated mixture population generated with fractions $\lambda=\{0.4,0.3,0.3\}$ is indicated with a black dashed line. The bottom panels show the observed distributions of BBHs in redshift and masses once an SNR cut of 12 is applied.}
    \label{fig:CE_distro}
\end{figure*}

\section{Case studies: the common envelope efficiency and progenitors metallicity}
\label{sec:5}

In this section, we present case studies to show how synthetic catalogs of BBH mergers can be used with GW population studies. We use the model by \citet{rahul} for our binary population. In this simple model, the binary population is based on the code \textsc{cosmic} \citep{Breivik:2019lmt}  to simulate BBH mergers. The star formation rate is parametrized in terms of star metallicity, galaxy mass, and redshift of formation of the binary following  \cite{2016MNRAS.463L..31L}.  For every value of metallicity, a population of BBH merger progenitors is generated by using \textsc{cosmic} to  evolve zero-age main-sequence stars, selecting those that form BBH mergers. The overall population of BBH mergers is obtained by re-weighting the cosmic BBHs progenitors by the star formation rate.

\subsection{\textsc{Cosmic} simulations}

\textsc{cosmic} simulates binary evolution based on prescriptions that model physical processes such as stellar winds, mass transfers between the binary, and supernovae kicks. One of the prescriptions of interest is the unstable mass transfer during the binary evolution that results in a common envelope (CE) phase parametrized by an efficiency $\alpha_{\rm CE}$.  Depending on the value of the CE efficiency, stellar binaries can be more or less efficient in producing BBH mergers \citep{2018MNRAS.477.4685B}. We explore the effect of CE efficiencies. Specifically, we consider CE efficiency values of $\alpha_{\rm CE} = \{0.3, 0.5, 1.0\}$.
The other prescriptions of the \textsc{cosmic} simulations are set to their default values reported on \textsc{cosmic} 3.4.0\footnote{\url{https://cosmic-popsynth.github.io/docs/stable/}}. As this study focuses on statistical inference, we choose not to optimize the model to fit observed distributions and rates. For each simulation, \textsc{cosmic} provides us with the distribution of time delays between the progenitor formation and the BBH merger. The procedure of building the population depends on the type of multi-channel analysis we consider (see later).

\subsection{Generation of the GW mock catalog}

To build a mock catalog of observed GW events, we use an approach similar to \citet{2018ApJ...863L..41F,2019ApJ...883L..42F} to simulate the detection of GW events and the estimation of source masses and redshift for each detected binary. For each binary, we calculate the matched filter SNR $\rho$ as
\begin{equation}
    \rho = 8 \left(\frac{\mathcal{M}_c}{26 \, M_\odot} \right)^{5/6} \left(\frac{d_L}{1500 \, {\rm Mpc}} \right) w,  
    \label{eq:snr}
\end{equation}
where $\mathcal{M}_c$ is the binary redshifted chirp mass and $d_L$ is the binary luminosity distance (calculated using a Planck cosmology \citep{2016A&A...594A..13P}). The scaling factors for the chirp mass and the luminosity distance are chosen to assume a network composed by LIGO Hanford, Livingston, and Virgo with typical detection ranges for O4 \citep{2018LRR....21....3A}. The luminosity distance scaling is calculated with the single-detector reach distances reported in \cite{2018LRR....21....3A}. The parameter $w$ is a scaling factor that takes into account the fact that not all the detectors in the network are optimally oriented with respect to the source position \citep{2015ApJ...806..263D}. The cumulative distribution of $w$ for a three-detector network is publicly available\footnote{\url{https://pages.jh.edu/eberti2/research/}}. 

Once the optimal SNR is calculated for each binary, we draw a ``observed'' SNR $\rho_{\rm obs}$ from a non-central $\chi^2$-square distribution (with non-centrality parameter $\rho$) and with a 6 degrees of freedom since we have 3 detectors in the network. Binaries are detected if they exceed an observed SNR of 12. For each detected binary, we then draw an ``observed'' chirp mass $\mathcal{M}_{c,{\rm obs}}$ and symmetric mass ratio $\eta_{\rm obs}$ using the same likelihoods in Appendix B of \citet{2019ApJ...883L..42F}. Once we are provided with a set of ``observed'' chirp masses, symmetric mass ratio and SNR, we generate mock posterior samples on their original ``true'' values using the likelihood models from which they were generated. These mock posterior samples are then converted to posterior samples in source frame masses and in redshift using Eq.~\eqref{eq:snr} (and correcting for the change of variable $\{\rho,\mathcal{M}_c,\eta\} \xrightarrow{} \{d_L,m_1,m_2\} $).

that we then use to generate mock posterior samples on redshift and source-frame masses for each detected signal.

We study the reconstruction of the BBH formation channels by extracting  64, 128, 256, 512, 1024, and 2048 binaries from the set of detected signals.
We use the statistical approach described in Sec.~\ref{sec:4} and the likelihood function in Eq.~\eqref{eq:pos} to find posterior distributions on the mixture coefficients. This study makes use of the \textsc{bilby} code \citep{2019ApJS..241...27A,2020MNRAS.499.3295R} and its nested sampling \textsc{dynesty} implementation \citep{2019S&C....29..891H} to sample from the posterior distribution of the mixture coefficients.

\subsection{Measuring the progenitors common envelope efficiency from BBHs ($\sum_j \lambda_j=1$)}
\label{sec:5.1}

In our first case study, we would like to infer CE efficiency from the observed GW events.  In this case, the formation channels from which we build our BBH catalogs are the \textsc{cosmic} simulations with $\alpha_{\rm CE}=\{0.3, 0.5, 1.0\}$. As described in Sec.~\ref{sec:2}, this is the case for which the progenitor population of BBHs is in common between the formation channels. The $\{\lambda\}$ represent the fraction of BBHs produced from progenitors with a given CE.
For our case study, we assume that $40\%$ of the BBH population is produced with $\alpha_{\rm CE}=0.3$, $30\%$ with $\alpha_{\rm CE}=0.5$ and $30\%$ with $\alpha_{\rm CE}=1.0$. Note that this is a toy model to show how the method can work, as massive binaries might not display different CE efficiencies \citep{2022arXiv220604062W}.

By using Eq.~\eqref{eq:overrate} and the chosen mixture coefficients, we build the overall BBHs merger rate. In Fig.~\ref{fig:CE_distro} we show the BBHs rate evolution in terms of redshift and masses for the three models and for the overall population that we simulate. While the three simulations predict similar mass distributions in shape, they significantly differ in terms of absolute merger rates \citep{ricker_timmes_taam_webbink_2018,2002MNRAS.329..897H,2022LRR....25....1M}.
We run the multi-channel analysis using two sets of priors for the fractional mixture coefficients. In the first analysis, we use a Dirichlet prior with concentration parameters fixed to $\{\zeta\}=\{0.5,0.5,0.5\}$ favoring single models, while in the second analysis we also allow the concentration parameters to vary in a uniform distribution between $[0.01,100]$.
\begin{figure}
    \centering
    \includegraphics{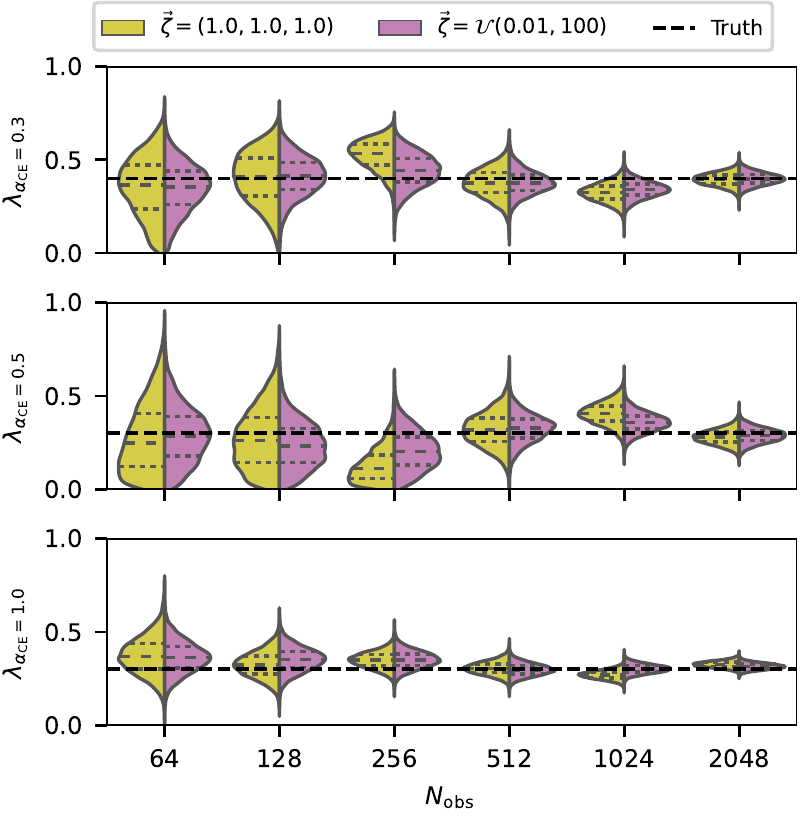}
    \caption{Posterior distributions on the fractional mixture coefficients for the three CE efficiency models as a function of the number of detected events. The horizontal black dashed line indicates the true value used for the simulation (that is always included in the 90\% credible intervals). The gray dashed lines in the posterior indicate the posteriors median and symmetric quartiles (50\% credible intervals). The yellow posteriors are generated by fixing a Dirichlet prior to distribution. The pink posteriors are generated allowing the Dirichlet parameter to change.}
    \label{fig:common_envelop_posterior}
\end{figure}
Fig.~\ref{fig:common_envelop_posterior} shows the reconstructed marginal posteriors of the mixture fractions between the three CE models. From the plot, we can see that when we are provided with few GW events, the constraints on the fractional mixture coefficients are weak. As more and more GW events are detected, the constraints on the mixture coefficients improve. With 2048 events a precision of $\sim 4-5\%$  is reached on the determination of the mixture fractions.  In App.~\ref{app:a} we also provide a more detailed discussion about the correlations among the various fractional coefficients and the Dirichlet concentration parameters.

\subsubsection{Reconstructing progenitors fractions from BBH fractions}
\label{sec:bias}
As we argue in Sec.~\ref{sec:2}, there is a fundamental difference between constructing the population probability using Eq.~\eqref{eq:res2} or Eq.~\eqref{eq:wrong}. In the former case, we are inferring the \textit{fraction of progenitors entering a formation channel} ($\lambda_j$), while in the latter we are inferring the \textit{fraction of BBHs produced in a formation channel} ($\Lambda_j$). These two quantities can be related a posteriori comparing Eq.~\eqref{eq:res2} and Eq.~\eqref{eq:wrong} and noting that
\begin{equation}
    \frac{\lambda_j N^{\varphi_{j}}_{\rm BBH}}{\sum_k \lambda_k N^{\varphi_{k}}_{\rm BBH}}=\Lambda_j.
\end{equation}
From the above relation, it follows that 
\begin{equation}
    \frac{\lambda_j}{\lambda_i }=\frac{\Lambda_j}{\Lambda_i}\frac{N^{\varphi_{i}}_{\rm BBH}}{N^{\varphi_{j}}_{\rm BBH}},
    \label{eq:rel}
\end{equation}
i.e. the ratio of the progenitors fraction entering the formation channel $j$ and $i$ can be calculated by scaling the ratio of BBH fractions produced in the formation channel $j$ and $i$ (and vice versa). Indeed it is interesting to note that the two ratios coincide when the formation channels have the same efficiency in producing BBHs.

\begin{figure}
    \centering
    \includegraphics[scale=0.5]{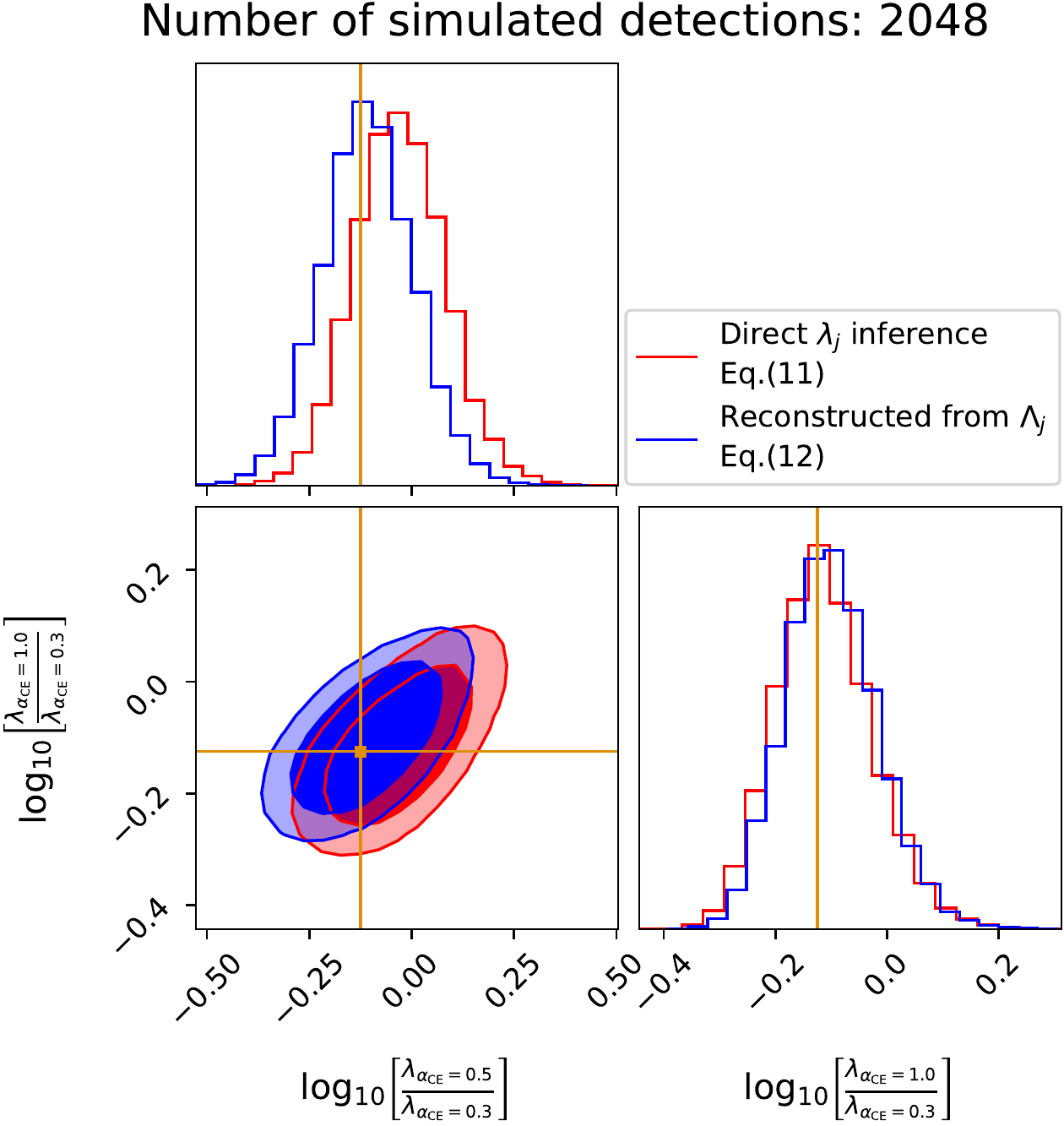}
    \caption{Joint posterior distribution on the progenitors $\lambda_j$ ratios for the CE simulations using 2048 GW detections. The figure compares the direct inference of the progenitor fractions (red contour) with the indirect reconstruction from BBHs population fractions (blue contour). The solid lines mark the $1\sigma$ and $2\sigma$ contours. The yellow solid lines mark the simulated population.}
    \label{fig:placeholder}
\end{figure}

For instance, in our previous example, we constructed a BBH population that was composed by 40\%, 30\%, and 30\% of progenitors with CE efficiency of 0.3, 0.5 and 1.0 respectively. We perform again the multi-channel analysis but this time using Eq.~\eqref{eq:wrong} and sampling for the $\Lambda_j$ (fraction of BBHs produced from the various CE efficiencies). 
In Fig.~\ref{fig:placeholder}, we compare the distribution of the progenitor ratios $\lambda_j/\lambda_i$ obtained in Sec.~\ref{sec:5.1} and the ones reconstructed using Eq.~\eqref{eq:rel} and the BBHs ratios $\Lambda_j/\Lambda_i$. We can see that the progenitors ratios can be effectively reconstructed from the BBHs ratios.

\subsection{Measuring the BBHs progenitors metallicity ($\sum \lambda_j \neq 1$)}

\
In the second case study, we use the simulation with CE efficiency $1.0$ and we divide the population of BBHs progenitors according to their metallicity, uniformly divided in base 10 logarithm between $\mathrm{Z}=5.0 \cdot 10^{-3} \mathrm{Z}_{\odot}$ and $\mathrm{Z}=1.6 \mathrm{Z}_{\odot}$. We are therefore in presence of independent sub-populations of BBHs progenitors, as described in Sec.~\ref{sec:2}. Each subpopulation of progenitors provides us with a sub-population of BBHs. The total BBH merger rate is the sum of the rates of these ten sub-populations. For this case study, we assume that the sub-populations are not present as predicted by the \textsc{cosmic} simulation. Instead, we assume that each subpopulation contributes to the overall BBHs merger rate with multiplicity coefficients $\lambda=\{0.5,0.6,0.7,0.8,0.9,1.0,1.1,1.2,1.3,1.4\}$ (ordered in terms of increasing metallicity bins). As an example, the BBHs produced from a progenitor with metallicity $\mathrm{Z}_\odot=2 \mathrm{Z}_\odot$ produce 40\% more BBHs with respect to the initial model predictions.
\begin{figure*}
    \centering
    \includegraphics{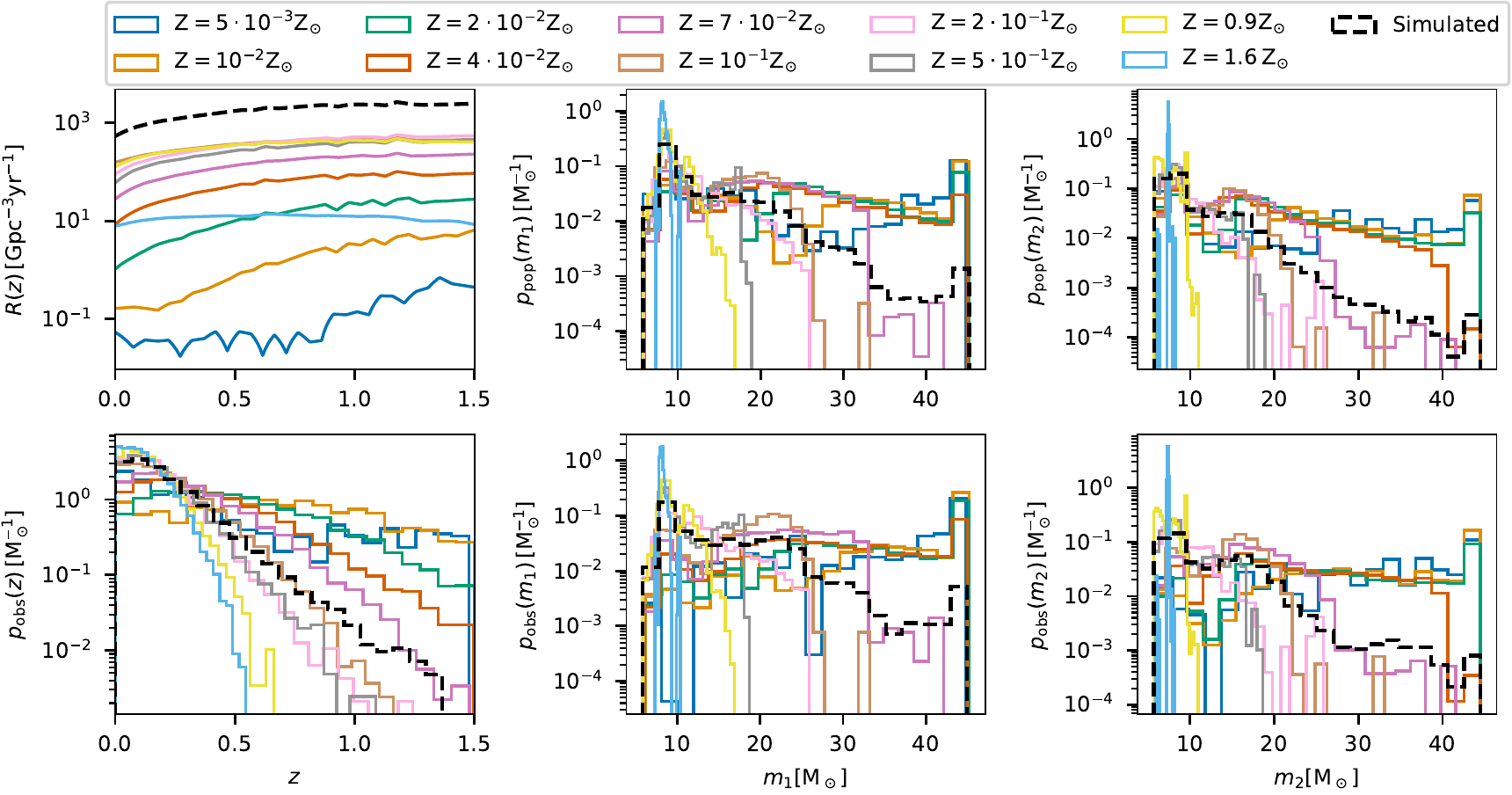}
    \caption{\textit{Left:} BBH merger rate redshift evolution for the metallicity subpopulations models considered. \textit{Center:} Distribution of the primary source mass for the models considered. \textit{Right:} Distribution of the secondary source mass for the models considered. The ``simulated'' curve shows the injected population and is obtained summing all the 10 metallicity bins sub-populations. The bottom panels show the observed distributions of BBHs in redshift and masses once an SNR cut of 12 is applied.}
    \label{fig:metal_distro}
\end{figure*}

Fig.~\ref{fig:metal_distro} shows the BBH merger rate and mass distributions for these sub-populations and the true population created from their sum. We  observe that in this simple model the progenitor metallicity introduces many different features  in the mass spectrum and merger rate.
\begin{figure*}
    \centering
    \includegraphics{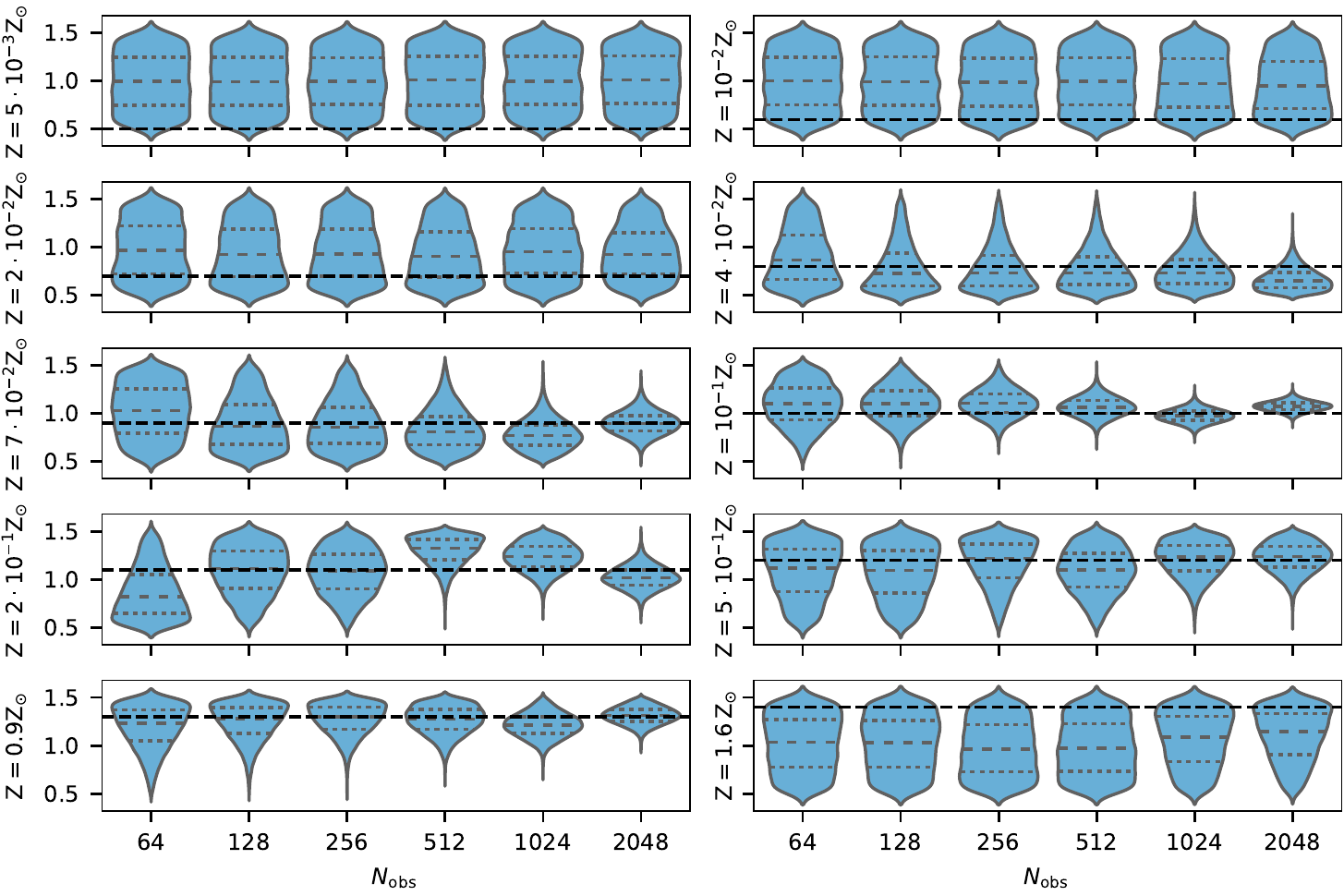}
    \caption{Posteriors of the mixture coefficients associated with the metallicity dependent subpopulations as a function of the number of  detected events. The horizontal black dashed line indicates the true value used for the simulation (that is always included in the 90\% CI). The gray dashed lines in the posterior indicate the posteriors' median and symmetric quartiles (50\% CI).}
    \label{fig:common_envelop_mets}
\end{figure*}
We perform the  reconstruction of the mixture coefficients for the sub-populations using priors on $\lambda_j$ independent from each other and uniform between 0.5 and 1.5. Fig.~\ref{fig:common_envelop_mets} shows the marginal posterior distribution among all the mixture coefficients. We can see that not all of the fractional parameters can be constrained in the proposed prior range. However, we can observe that in general,  models which predict more BBHs are better constrained than models that predict less. This is expected as progenitors entering channels predicting more BBHs are easier to constrain. This case study shows that synthetic binary catalogs can be used with GW events to probe the BHs progenitor metallicity but also the evolution of the star formation rate.

\newpage

\section{Conclusions}

In this paper, we have described in detail how to employ and statistically interpret synthetic compact binaries in light of GW detections. In particular, we have focused on analyses trying to infer and constrain the presence of BBHs progenitors in multiple channels.

In Sec.~\ref{sec:3}, we have presented for the first time an efficient method to evaluate the ``match'' between synthetic catalogs of binaries and phenomenological reconstructed astrophysical rates. Given the phenomenological rate reconstruction, the method is able to assign a probability to each of the synthetic catalogs to be representative of the estimated rate. The probabilities can be used to quickly evaluate how much a model fits the phenomenological rates with respect to the other.

In Sec.~\ref{sec:2}, we have formalized how different progenitor populations can be used to build multi-channel population models. We have discussed how an overall BBHs merger rate should be built and interpreted in terms of progenitors mixture coefficients $\{\lambda\}$. We have argued that in the case that the progenitor population is common across the different BBHs formation channels, then one can use fractional mixture coefficients to infer the percentage of progenitors undergoing trough to each formation channel ($\sum_j \lambda_j=1$). We have also discussed the case for which we are in presence of multiple and independent sub-populations, showing that in this case the $\{\lambda\}$ can be assumed to be independent of each other. From an astrophysical perspective, normalized mixture coefficients can be used when building multi-channel progenitors models for which only the stellar evolution is modified. While independent mixture coefficients can be used when stellar evolution models are fixed, but the original progenitor rates (initial conditions) are varied.

In Sec.~\ref{sec:4} we have reviewed the hierarchical statistical method used to employ synthetic populations with observed GW events, describing the method in light of the multi-channel analyses presented in Sec.~\ref{sec:2}. In Sec.~\ref{sec:4} we have also described technical aspects related to the computational implementation of this methodology and the prior choice that should be made on the mixture coefficients in order to not introduce an ``ordering preference''. 

In Sec.~\ref{sec:5} we have presented two case studies for BBH progenitors' multi-channel analyses. In the first case, we discussed a possible measure of the CE efficiency parameter. Based on an astrophysical model for BBH formation, we show that binary evolution criteria, such as the CE efficiency, could be constrained to good precision with a few thousand of detection (or in the coming years). The second case that we discussed, made use of BBH sub-populations divided into progenitor metallicity bins. We have shown that, provided the astrophysical model and star formation rate, some of the BBHs progenitors' metallicity can be constrained with thousands of GW detections.

With the next two observing runs O4 and O5, the LIGO, Virgo, and KAGRA detectors will reveal thousands of BBHs and possibly hundreds of BNSs \citep{2018LRR....21....3A}. Using this observed population it will be possible to probe the progenitor properties of the GW sources and unveil the astrophysical processes bringing to compact object formation.

\label{sec:6}

\section*{Acknowledgements}

SM, AL and TB are supported by the ANR COSMERGE project, grant ANR-20-CE31-001 of the French Agence Nationale de la Recherche. This work was supported by the "Programme National des Hautes Energies" (PNHE) of CNRS/INSU co-funded by CEA and CNES" and the authors acknowledge HPC ressources from "Mesocentre SIGAMM" hosted by Observatoire de la Côte d’Azur. We thank M.~Zevin, C. Berry, R.~O'Shaughnessy and O.~J.~Piccinni for comments and discussion during the internal LIGO-Virgo circulation of this work.
.
\section*{Data Availability}

Results and simulations obtained in this paper are generated using the \textsc{asimovgw} python package, that can be download at \url{https://github.com/simone-mastrogiovanni/gwparents}.



\bibliographystyle{mnras}
\bibliography{example} 




\appendix
\onecolumn
\section{Common envelope efficiency with Dirichlet concentration parameters}
\label{app:a}
\begin{figure*}
    \centering
    \includegraphics[scale=0.4]{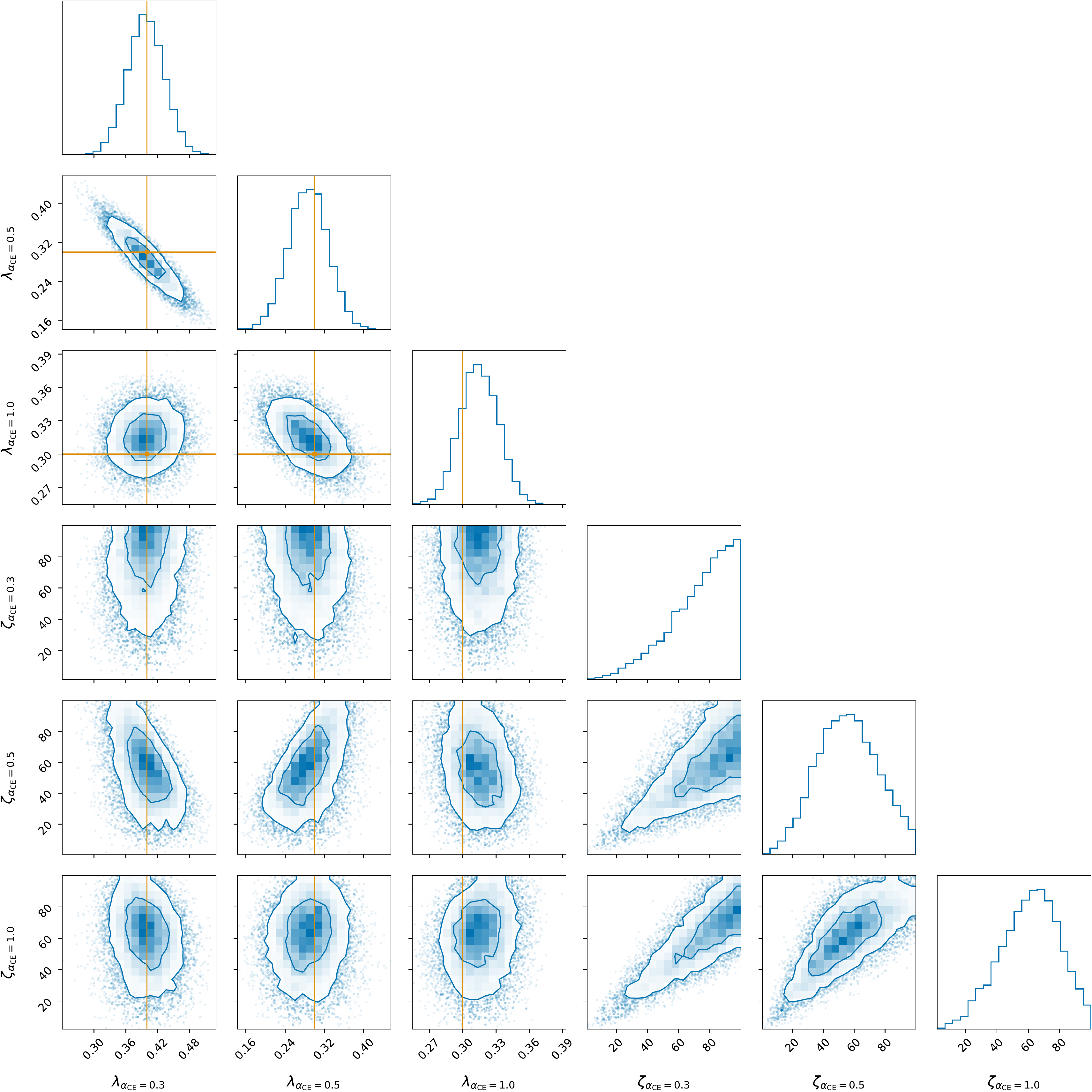}
    \caption{Corner plots of the posterior on the fractional mixture coefficients and concentration parameters of the Dirichlet distribution obtained for the CE efficiency run and 2048 GW detections. The blue solid lines indicate the injected values for the fractional mixture coefficients.}
    \label{fig:alpha_corners}
\end{figure*}

In this appendix, we run the same inference on the progenitor fractions as in Sec.~\ref{sec:5.1} but using also priors on the Dirichlet concentration parameters. 
It is interesting to see what are the correlations in the determination of the mixture fractions and the concentration parameters $\alpha$. In Fig.~\ref{fig:alpha_corners}, we show their joint posterior distribution for 2048 GW detections. We note that the fractional mixture parameters of the CE efficiency $0.5, 1.0$ show a non-negligible anti-correlation.
This is due to the fact that these two formation channels predict similar values (and higher with respect to $\alpha_{\rm CE}=0.3$) of the BBH merger rate, see Fig.\ref{fig:CE_distro}. These two models are anti-correlated as they cannot both be present with high fractions, otherwise, they would overestimate the overall merger rate. On the other hand,  one can see that the coefficient corresponding to $\alpha_{\rm CE}=0.3$ does not show any significant correlation with the others. This is due to the fact that the CE population has a negligible BBH merger rate if compared to the other two. The concentration parameter $\{\zeta\}$ acts as a ``nuisance'' parameter for determining the prior weights on the mixture coefficients. It is interesting to note however that all the concentration parameters are correlated. This is due to the fact that, given a Dirichlet distribution on $\lambda_j$ parameters with concentration parameters $\zeta_j$, the expected values of $E[\lambda_j]=\zeta_j/\sum_k \zeta_k$.

\bsp	
\label{lastpage}
\newpage
\clearpage
\end{document}